\documentclass{llncs}
\usepackage{amsmath,amssymb}
\usepackage{graphicx}

\begin{document}
\title{Approximation Algorithms for the Asymmetric Traveling Salesman Problem : Describing two recent methods}

\subtitle{}

\author{Arka Bhattacharya}

\institute{Columbia University,\ New York\ \email{\it (ab3899@columbia.edu)}}

\maketitle

\begin{abstract}
The paper provides a description of the two recent approximation algorithms for the Asymmetric Traveling Salesman Problem, giving the intuitive description of the works of Feige-Singh[1] and Asadpour et.al\ [2].\newline 
[1] improves the previous $O(\log n)$ approximation algorithm, by improving the constant from 0.84 to 0.66 and modifying the work of Kaplan et. al\ [3] and also shows an efficient reduction from ATSPP to ATSP. Combining both the results, they finally establish an approximation ratio of $\left(\frac{4}{3}+\epsilon \right)\log n$ for ATSPP,\ considering a small $\epsilon>0$,\  improving the work of Chekuri and Pal.[4]\newline 
Asadpour et.al, in their seminal work\ [2], gives an $O\left(\frac{\log n}{\log \log n}\right)$ randomized algorithm for the ATSP, by symmetrizing and modifying the solution of the Held-Karp relaxation problem and then proving an exponential family distribution for probabilistically constructing a maximum entropy spanning tree from a spanning tree polytope and then finally defining the thin-ness property and transforming a thin spanning tree into an Eulerian walk.\ The optimization methods used in\ [2] are quite elegant and the approximation ratio could further be improved, by manipulating the thin-ness of the cuts.  
\end{abstract}

\section{Introduction and Basic Definitions}
We give some basic definitions in this section. \newline
Considered by many, to be the most famous NP-Complete problem, The Traveling Salesman Problem (TSP) takes any graph, $G=(V,E)$ with edge weights, as an input and outputs a minimum length tour, which spans all the vertices, such that each vertex, $v\in V$ appears exactly once on the tour.\ Consider that we have an edge weight function, $w:E\times E\rightarrow R$, such that it satisfies the triangle inequality, $w(u,v)+w(v,w)\geq w(u,w)$ for some $u,v\in V$. The metric version takes a complete graph and the weight function as an input and it is known that a metric version of the problem always has a cyclic tour, while a non-metric version may not have a tour at all. So, we are generally interested in finding poly time approximation schemes for the metric TSP.\newline
The symmetric version has edge weights exactly the same in both the directions of the edge from two vertices and hence, does not apply to directed graphs while the asymmetric problem (ATSP) deals with graphs, where the anti-parallel edges do not have the same weights. It is easy to see that the symmetric version is a spacial case of the asymmetric one. We will consider the metric ATSP for our analysis.\newline
Section 2 gives the total analysis of the Feige-Singh algorithm and Section 3 gives the analysis of the algorithm of Asadpour et.al.

\section{The Feige-Singh Algorithm}
[1] provides a modest improvement in the leading constant of the approximation ratio, which was achieved by [3] and the total result is summarized in the following theorem below.\newline \newline
\textit{Theorem 1:} 
\ There exists a polynomial time algorithm which returns a Hamiltonian cycle of weight at most $\frac{2}{3}\log n$ of the weight of the minimum cost Hamltonian cycle in a given directed graph, $G=(V,E)$, having a weight function, $w$ satisfying the triangle inequality.\newline \newline
Let us look at some of the definitions, that we will be using throughout.\ Consider that, we are given a graph $G=(V,E)$, a starting vertex $s$ and a terminal vertex $t$.\newline
\textit{Definition 2.1:}\ We call a $(s,t)$ walk in $G$ \textit{\textbf{spanning}} if it visits every vertex of $G$ at least once and thus, vertices and edges can appear more than once in the walk.\ A tour is an $(s,s)$ walk which is spanning.\newline
\textit{Definition 2.2:}\ Given a directed path $P$ and vertices $u$ and $v$ on $P$, such that $v$ occurs after $u$ on $P$, we denote $P(u,v)$ to be the \textit{\textbf{subpath of P}}, starting at $u$ and ending at $v$.\newline
\textit{Definition 2.3:}\ Given two paths, $P$ and $Q$, we say that \textit{\textbf{Q respects the ordering of P}} if $Q$ contains all the vertices of $P$ and for every two vertices $u$ and $v$ in $P$, $u$ appears before $v$ in $Q$ iff $u$ appears before $v$ in $P$. \newline \newline
The Asymmetric Traveling Salesman Path Problem (ATSPP) is an interesting variant of the ATSP, in which we need to find a minimum weight Hamiltonian Path from nodes $s$ to $t$, given an instance of the graph, the weight function $w$, satisfying triangle inequality and the two nodes, $s$ and $t$.It will be shown in the next subsections, that the ATSPP can be approximated nearly as well as the ATSP.\newline
Say, we have the starting node $s$ and the ending node, $t$ as an instance to the ATSPP and let $OPT$ denote the spanning path from $s$ to $t$ of minimum cost. We assume WLOG, that for every two vertices $u$ and $v$ in the graph, there  exists an edge $(u,v)$ which is the shortest distance between them.\newline 
There is a spanning minimum weight non-simple path between $s$ and $t$ and if we have the value of $OPT$ lesser than the distance between $t$ and $s$ (denoted by $d(t,s)$), we simply remove all incoming and outgoing edges from $s$ and $t$ respectively(hence, we formally made them the starting and ending nodes) and update the edge weight of the newly constructed edge $(t,s)$, by $min\{OPT,d(t,s)\}$ and thus, we get a non-simple ATSP tour of weight $(d(t,s)+OPT)$, which can be at most $2OPT$. Assuming that we have the $\alpha$-approximation algorithm for the ATSP, we find a simple ATSP tour of weight not exceeding $2\alpha OPT$ and remove the edge $(t,s)$, thus having a spanning path from $s$ to $t$ of weight at most $2\alpha OPT-OPT=(2\alpha -1)OPT$. But, if the weight of the edge $(u,v)$ becomes smaller than the original graph when we added the edge $(t,s)$, then we add the path $u-t-s-v$ and hence, the edge $(t,s)$ re-appears again. Hence, if we do this for $r$ times, $(t,s)$ gets added $r$ times and we have to remove it from all the $r$ cases to get $r$ spanning paths covering all the vertices together as a whole, such that the summation of all the path weights is at most $(2\alpha -r)OPT$.\ Now, we show in lemma 1, that such a single path from $s$ to $t$ exists, which respects the order of each of the paths and does not have much increase in its weight.\newline \newline 
Now, we construct an algorithm for the ATSPP, given an $\alpha$ approximation algorithm for the ATSP.\newline
Let us have an instance, $I=(G,w,s,t,\epsilon,A)$ for the ATSPP, where $A$ is the $\alpha$-approximation algorithm for the ATSP, $\epsilon > 0$ is some parameter and $w$ satisfies the triangle inequality for the directed graph $G$, having $n$ vertices.\ Consider the algorithm, $B$ below for ATSPP.\newline \newline
$B(G,w,s,t,\epsilon,A)$ \newline \newline 
1. Find some positive $d\in \left[\left(1-\frac{\epsilon}{8} \right)OPT,OPT \right]$.\newline 
2. Construct graph $G_{1}$ by removing incident edges on $s$ and outgoing edges from $t$ and adding an edge $(t,s)$ of weight $d$.Let the weight function of this graph be $w_{1}$.\newline 
3. Call $A$ on the complete directed graph (say $G_{2}$) on the set of vertices of $G$ and let $w_{2}$ be the edge weights of $G_{2}$.Let the $\alpha$-approximate solution be $S$.\ ($w_{2}(u,v)$ is nothing but the shortest distance from $u$ to $v$ under $w_{1}$)\newline 
4. Replace each edge $(u,v)$ in $G_{1}$ by its corresponding shortest path and obtain a tour, say $T$ in $G_{1}$.\newline 
5. Decompose $T$ into a collection of $r$ subpaths,\ $P$ spanning all the vertices,\ where $r$ is the no. of times the edge $(t,s)$ appears and then shortcut them to a single path, where each vertex (except of course $s$ and $t$) appears exactly once in one of the paths.\newline 
6. Return $C(P,\epsilon)$.\newline \newline 
$C$ is the algorithm, which actually tries to construct a single path, out of all the $r$ subpaths, such that the weight of the resulting path is not very large and follows the order of all the original paths.\ It takes the collection $P=\{P_{1},P_{2},...,P_{r}\}$ of $(s,t)$-paths and $\epsilon$ as its input.\newline \newline \newline \newline
$C(P,\epsilon)$ \newline \newline 
1. If $r=1$ return $P_{1}$, else let $k=min\{\frac{9}{\epsilon},r\}$ \newline 
2. Find minimum weight path $P^{,}$, spanning all the vertices of $(P_{1},P_{2},...,P_{k})$, which respects their order.\newline 
3. $P^{,}=P\cup (P^{,}\backslash \{P_{1},P_{2},...,P_{k}\})$ \newline 
4. return $C(P^{,},\epsilon)$ \newline
The construction and performance of the algorithms described above, which help in the conversion of ATSP to ATSPP are based on the lemma described next. \newline \newline 
\textit{Lemma 1:}\ Consider we have a collection of $k$ paths $P_{1},P_{2},...,P_{k}$ from $s$ to $t$, such that no vertex appears in more than one path, then we can construct a single path from $s$ to $t$ that spans all the vertices in the $k$ paths, respects the order of each of the original paths and weighs no more than the sum of the summation of the weights of all the $k$ paths and $k.OPT$, where $OPT$ is the weight of the minimum ATSPP.\newline \newline 
Though, we won't give the whole proof, but would mention the main points, on which the proof stands. The proof is simple and the reader can refer [1] for the details. \newline
For each path, $P_{i}$ we create prefix paths $Q_{i}$ and a path, $Q$, which is to be constructed from $s$ to $t$ in each iteration, such that it covers all the vertices of the prefix paths and maintain their order. We maintain nodes, known as the front nodes for each path $P_{i}$ and it is nothing but the successor of each $Q_{i}$ in $P_{i}$, when the algorithm runs iteratively. In simple words, $front_{i}$ for each path $P_{i}$ is the vertex,such that all the ordered vertices before it in the path have already been explored by the algorithm. Let us assume that the optimal ATSPP from $s$ to $t$ is the path $P$ and $P(u,v)$ is the subpath from any two vertices $u$ and $v$ in the path $P$.Let $v$ be the last vertex of the path $Q$. \textit{The main argument of the proof is an invariant, that all the front vertices for each path, $P_{i}$ occur in the subpath, $P(v,t)$.(The front vertex of some path $P_{j}$, which contains $v$ may or may not be in the subpath $P(v,t)$)}. So, if we initialize the path $Q$ and all the prefix paths to be $(s)$ and the front vertex for each path, to be the second vertex for each given $P_{i}$, then the invariant gets trivially satisfied. In every iterative step of the algorithm, one or two paths advance their front vertex and $Q$ continues to be created till it reaches the terminal vertex, $t$. [1] also shows that any edge of $P$ can be used at most $k$ times.Moreover, the subpaths of $P_{i}$ in the path, $Q$ are edge disjoint for all $i$, which can be easily deduced from the statement of the invariant and the functioning of the iterative steps of the algorithm. Hence, the weight of the single path, which we constructed from $s$ to $t$, can never exceed the sum of the summation of the weights of all the paths, $P_{i}$ and $k$ times the weight of the optimal path, $P$. Thus, $w_{1}(Q)\leq \sum_{i=1}^{k}w_{1}(P_{i})+k.w_{1}(P)=\sum_{i=1}^{k}w_{1}(P_{i})+k.OPT$.\newline \newline 
Now, by simple observation, if we replace all the $P_{i}^{,}$s by $Q$, then in each iteration, the weight of the new collection of paths increase by $k.OPT$, by Lemma 1 and number of paths reduce by $k-1$. The maximum number of iterations of algorithm $C$ could be $\lfloor(\frac{r}{\left(\frac{9}{\epsilon}-1\right)}\rfloor\leq 1+\frac{\epsilon r}{8}$ and hence, the maximum increase in weight could be $\left(r-1+\frac{\epsilon r}{8}+1\right)=\left(1+\frac{\epsilon}{8}\right).r.OPT$, for a collection of $r$ $(s,t)$ paths. Thus, for the final constructed path, $Q$ by the algorithm, we have $w_{1}(Q)\leq \sum_{i=1}^{r}w_{1}(P_{i})+\left(1+\frac{\epsilon}{8}\right).r.OPT$.\newline \newline
We know that $OPT$ is the weight of the optimal spanning path from $s$ to $t$ in $G_{1}$, under the weight function $w_{1}$.Since, the weight of the edge $(t,s)$ in $G_{2}$ is $d$, hence the optimal Hamiltonian path in $G_{2}$ has a weight of $(OPT+d)$.If we remember the functioning of algorithm $B$, we called $A$ on $G_{2}$ and hence, returns a Hamiltonian cycle, say $C$ of $G_{2}$, of a weight not exceeding $\alpha (OPT+d)\leq 2\alpha OPT$.\ Now, if we remove all the $r$ copies of the edge $(t,s)$, having a weight $d$ and decompose the path, $T$ into a single path spanning all the vertices, then the summation of the weights of all the initial $r$ $(s,t)$ paths gets decreased by $rd$ from the previous value and hence, $\sum_{i=1}^{r}w_{1}(P_{i})\leq 2\alpha OPT-rd$.\ Also, algorithm $C$ returns a single path, $Q$ such that $w_{1}(Q)\leq \sum_{i=1}^{r}w_{1}(P_{i})+\left(1+\frac{\epsilon}{8}\right).r.OPT$ in time $O(n^{k})=O(n^{\frac{1}{\epsilon}})$. Thus, $w_{1}(Q)\leq 2\alpha OPT-rd+\left(1+\frac{\epsilon}{8}\right).r.OPT$.\ Using, $d\geq \left(1-\frac{\epsilon}{8}\right).OPT$ and $rd\leq \alpha(OPT+d)$, we get $w_{1}(Q)\leq (2+\epsilon)\alpha OPT$.\ Now, we are in a position to construct a theorem and is written below.\newline \newline 
\textit{Theorem 2}:\ Consider that, we are provided with a directed graph, $G$ with a weight function $w$, satisfying triangle equality, vertices $s$ and $t$ and an $\alpha$ approximation algorithm to the ATSP, such that $OPT$ is the minimum weight Hamiltonian path from $s$ to $t$. Then, there exists an algorithm, which gives a Hamiltonian path from $s$ to $t$ of weight no more than $(2+\epsilon)\alpha OPT$, for some constant positive $\epsilon$. \newline \newline 
Finally, they showed that the KLSS algorithm has not tight bounds and could be improved to $0.787\log n$ approximation and ending by further improving the approximation ratio to $0.667\log n$, by slightly modifying the algorithm.\ The KLSS analysis, uses the following LP for ATSP, which enforces sub-tour elimination constraints for subsets of size two. \newline \newline 
minimize $\sum_{e\in E}c_{e}x_{e}$ \newline \newline 
s.t\ \ $\sum_{e\in \delta^{+}(v)}x_{e}=\sum_{e\in \delta^{-}(v)}x_{e}=1 \ \ \ \forall v\in V$ \newline \newline 
$x_{(u,v)}+x_{(v,u)}\leq 1\ \ \forall u,v\in V$ \newline \newline 
$0\leq x_{e}\leq 1\ \ \forall e\in E$ \newline \newline 
$\delta^{+}(v)$ and $\delta^{-}(v)$ are the set of edges going out and coming in, of the vertex $v$ and $x_{e}$ for an edge $e$ is set to 1, if it is in the Hamiltonian cycle.\ The main lemma, on which the KLSS construction depends, is stated below.\newline \newline 
\textit{Lemma 2}:\ Consider, we are given a directed graph, $G$ whose edges are weighted and let us have an optimal solution,$OPT$,\ to the LP-ATSP problem, described above.\ Then, there exists an efficient algorithm which finds two cycle covers, $C_{1}$ and $C_{2}$ such that they do not have a common 2-cycle and the summation of their weights, does not exceed $2.OPT$. \newline \newline 
KLSS algorithm finds such a $C_{1}$ and $C_{2}$ and constructs another cover, $C_{3}=C_{1}\cup C_{2}$ and chooses only that cover, such that it satisfies, $min \ \left\{\frac{w(C_{1})}{\log \frac{n_{i}}{c(C_{1})}},\frac{w(C_{2})}{\log \frac{n_{i}}{c(C_{2})}},\frac{w(C_{3})}{\log \frac{n_{i}}{c(C_{3})}}\right\}$, where $n_{i}$ is the number of nodes for the $i^{th}$ iteration and $c(.)$ gives the number of components, for the cover.\ Finally, for each connected component, we pick one representative vertex and delete the rest of them and continue the iteration to the next stage, until we have only one component left.\ We simply return the solution, $Cover_{1}\cup Cover_{2}\cup ...\cup Cover_{k}$ ,where $Cover_{i}$ denotes the cover selected in the $i^{th}$ iteration and $k$ is the total number of iterations. \newline
The improvement in the approximation ratio is achieved by scaling down the values of $w(C_{i})$ and $c(C_{i})$, such that we have new variables, $w_{i}=\frac{w(C_{i})}{OPT}$ and $c_{i}=\frac{c(C_{i})}{n_{i}}$ and formulating a new LP problem on these variables and also, adding all the previous LP constraints as a representation of these variables.\ (Refer [1] for the algebraic proof.) \newline \newline 
At the end, KLSS algorithm is slightly modified, where instead of choosing one of the covers, $C_{1}$, $C_{2}$ or $C_{3}$ minimizing the potential function, $\frac{w(.)}{\log \frac{n_{i}}{c(.)}}$, we decompose $C_{3}$ into two Eulerian sub-graphs.\ The construction relies on the main lemma below. \newline \newline 
\textit{Lemma 3}:\ Consider, we have a directed graph, $G=(V,E)$,where $|V|\geq 3$, which is connected and does not contain self-loops, such that every vertex has both in and out degrees to be two.\ Then, there exists either two vertex disjoint cycles of length 2 or one cycle of length at least 3, such that if we remove all the edges of these cycles from the graph, the graph still remains connected. \newline \newline 
Now, initially we had three cycle covers and we chose, the best of them in each iteration. But, in this case of modified version, we create $C_{4}$ and $C_{5}$, where $C_{5}$ is the set of cycles, chosen from each component of $C_{3}$ (each of the components remains connected by Lemma 3) and $C_{4}=C_{3}\backslash C_{5}$.\ In this case, for each iteration, we choose either $C_{4}$ or $C_{5}$, whoever minimizes the function, $\frac{w(.)}{\log \frac{n_{i}}{c(.)}}$, where $n_{i}$ is the number of nodes in the current graph.\ The remaining algorithm stays the same as the previous.\newline 
Finally, using the scaled down version of the functions, $w(.)$ and $c(.)$ and slightly modifying the LP relaxation problem, we obtain that for each iteration,$i$, $\frac{w(Cover_{i})}{\log \frac{n_{i}}{c(Cover_{i})}}\leq \frac{2}{3}.\log n.OPT$, where $Cover_{i}$ must be either $C_{4}$ or $C_{5}$ and thus, giving a deterministic $O(\log n)$ approximation algorithm for the ATSP, with an improved leading constant.

\section{The Asadpour-Goemans-Madry-Gharan-Saberi randomized approximation algorithm}
[2] provides an $O\left(\frac{\log\ n}{\log\ \log n}\right)$ randomized algorithm for the asymmetric TSP, for edge weights satisfying the triangle inequality.\ Let us describe some notations.\ We use $(u,v)$ to denote a directed edge from $u$ to $v$, while $\{u,v\}$ for an undirected edge and $A$ and $E$ to be the set of directed and undirected edges in a directed and undirected graph respectively.\ The cost of a function,\ $f:A\rightarrow R$,denoted by $c(f)$, can be defined as $c(f):=\sum_{a\in A}c(a)f(a)$ and consider that we have a subset, $S$ of $A$, then we have $f(S)=\sum_{a\in S}f(a)$.\ For a directed graph, with the vertex set as $V$, consider the following notations below for $U\subseteq V$ :\newline \newline 
$\delta^{+}(U):=\ \{a=(u,v)\ s.t \ u\in U,v\not\in U, (u,v)\in A\}$ \newline 
$\delta^{-}(U):=\ \{a=(u,v)\ s.t \ u\not\in U,v\in U, (u,v)\in A\}$ \newline 
$A(U):=\ \{a=(u,v)\ s.t \ u\in U, v\in U\, (u,v)\in A\}$ \newline \newline 
Now, similar notations can be written for an undirected graph, with edge set $E$ and the same vertex set, $V$ as below for $U\subseteq V$ :\newline \newline
$\delta(U):=\{e=\{u,v\}\ s.t \ u\in U,v\not\in U\ or\ u\not\in U,v\in U, \{u,v\}\in E\}$ \newline 
$E(U):=\{e=\{u,v\}\ s.t\ u\in U,v\in U,\{u,v\}\in E\}$ \newline \newline 
Now, let us discuss the Held-Karp relaxation and obtain, a scaled down symmetric solution lying within a convex spanning tree polytope.\newline 
Before defining the relaxation, let us define the variable $x_{a}$, for an edge $a=(u,v)$. We have, $x_{a}=1$, if the edge $a$ is in the optimal tour, otherwise it is $0$. We relax this variable,from $\{0,1\}$ to $[0,1]$.\ Consider, we are given an instance of the ATSP, such that the cost function, $c:V\times V\rightarrow R^{+}$ is well defined, then a lower bound could be obtained on the optimal tour cost if we solve the following LP below, defined on the complete bi-directed graph with $V$ as the vertex set : \newline \newline 
\ \ \ \ \ minimize $\sum_{a}c(a)x_{a}$ \newline \newline
\ \ \ \ \ s.t \ $x(\delta^{+}(U))+x(\delta^{-}(U))=\sum_{a\in \delta^{+}(U)}x_{a}+\sum_{a\in \delta^{-}(U)}x_{a}\geq 2$ \ \ \ \ $\forall U\subset V$ \newline \newline 
$x(\delta^{+}(v))+x(\delta^{-}(v))=2$ \ \ \ \ $\forall v \in V$ \newline \newline 
$0\leq x_{a}\leq 1$\ \ \ \ $\forall a$ \newline \newline 
Let the optimal tour solution of the LP be $x^{*}$, such that the cost of this tour is $OPT_{HK}$, i.e $c(x^{*})=OPT_{HK}$ and hence, we could assume that $x^{*}$ is an extreme point of the corresponding polytope. For, an undirected edge, ${u,v}$, we define a symmetric scaled down version of the Held-Karp solution, so that the resuting vector supports Edmond's characterization of the base polytope of a matroid and thus, can be made to fall in the relative interior of a spanning tree polytope.(refer [2] for the short proof). We have,
\begin{equation*}
z^{*}_{\{u,v\}}:=\frac{n-1}{n}(x^{*}_{uv}+x^{*}_{vu}) 
\end{equation*}
Thus, we have,
\begin{equation*}
z^{*}\delta(U)\geq 2\left(1-\frac{1}{n}\right) 
\end{equation*} 
Let, $A$ and $E$ denote the support of the vectors $x^{*}$ and $z^{*}$ respectively and let the cost, $c(e)$ of an undirected edge, $e=\{u,v\}$ be the minimum of the cost of the two directed edges, $(u,v)$ and $(v,u)$ common to the support, $A$. Hence, it is easy to see that $c(z^{*})<c(x^{*})$.\ Hence, we could see that the vector, $z^{*}$ can be expressed as a convex combination of the spanning trees, such that the co-efficient corresponding to every spanning tree is positive.\ Now, the next goal is to round $z^{*}$ from a point in the relative interior of the spanning tree polytope to a spanning tree, using a distribution over the spanning trees of the graph, $G$, such that the marginal probability imposed by $z^{*}$ is preserved.\newline 
Consider, we have a collection of all the spanning trees, $T$ of the graph, $G=(V,E)$ and we denote this set, by $\tau$. Let the entropy distribution function function be represented by $p(.)$. Then, the maximum entropy distribution, say $p^{*}(.)$, with respect to given marginal probabilities, say $z$ is known to be the optimum solution of the following convex program\ (T denotes a spanning tree): \newline \newline 
infimum \ $\sum_{T\in \tau}p(T)\log p(T)$ \newline \newline 
s.t \ \ $\sum_{T\ni e}p(T)=z_{e}$\ \ \ $\forall e\in E$ \newline \newline 
$p(T)\geq 0$ \newline \newline 
Let us construct a Lagrange function, $L(p,\delta)$, by associating a Lagrange multiplier, $\delta_{e}$ for every edge $e\in E$, for which the marginal probability is $z_{e}$ and we have, $\sum_{e\in T}\delta_{e}=\delta$. We have,
\begin{equation*}
L(p,\delta)=\sum_{T\in \tau}p(T)\log p(T)-\sum_{e\in E}\delta_{e}\left(\sum_{T\ni e}p(T)-z_{e}\right)
\end{equation*}
\begin{equation*}
= \sum_{e\in E}\delta_{e}z_{e}+\sum_{T\in \tau}\left(p(T)\log p(T)-\delta p(T) \right)
\end{equation*}
Hence, for the Lagrange dual, we have to calculate the supremum of the infimum of the Lagrange function, i.e $sup_{\delta}inf_{p\geq 0}L(p,\delta)$. For the infimum, we could actually minimize the function, $(p(T)\log p(T)-\delta p(T)$, use basic differential calculus of analyzing the maxima/minima and then find the function, $p(T)$. The function, $p(T)$ is found to be an exponential function after the calculations. We have $p(T)=e^{\gamma (T)}$, where $\gamma(T)=\delta(T)-1$. (Thus, for each edge we have $\gamma_{e}=\delta_{e}-\frac{1}{n-1}$)\ The value of the infimum becomes, $(1+\sum_{e\in E}z_{e}\gamma_{e}-\sum_{T\in \tau}e^{\gamma(T)})$.\ Hence, for the Lagrange dual, we need to find the supremum of this infimumn, for all values of $\gamma$. Let, the vector, which makes the value of this dual, equal to the optimum value of the original convex program, be $\gamma^{*}$.\ By the analysis of saddle points, we can say that $p^{*}(T)$ is the unique minimizer of the Lagrange dual, $L(p,\gamma^{*})$. Hence, we have $p^{*}(T)=e^{\gamma^{*}(T)}$.\ So, we are in a position to write the first theorem, after the construction of the exponential distribution, for sampling the spanning trees, which preserves the values of the marginal probabilities of the edges.\newline \newline 
\textit{Theorem 1}:\ There exist $\gamma_{e}^{*}$ for all edges, $e$, such that if we sample a spanning tree, $T$ of the graph, $G$ according to the exponential distribution, $p^{*}(T)=e^{\gamma^{*}(T)}$, then the marginal probability of an edge belonging to the tour is preserved with respect to a given vector, which lies in the relative interior of a spanning tree polytope. \newline \newline 
Let us define an exponential family distribution,$\tilde{p}(T)$.\ Consider, for a collection of spanning trees, $\tau$, we define $\tilde{p}(T)=\frac{e^{\tilde{\gamma}(T)}}{\sum_{T\in \tau}e^{\tilde{\gamma}(T)}}$. After, constructing this distribution, a combinatorial approach for efficiently calculating the values of the $\tilde{\gamma_{e}}^{,}$s for all edges, $e\in E$ can be developed, such that for a given $z$ in the spanning tree polytope of the graph, $G=(V,E)$ and some $\epsilon >0$, if we sample a spanning tree according to the distribution, the value of $\tilde{z_{e}}$ for each edge, never exceeds $(1+\epsilon)z_{e}$ and hence, the marginals are preserved. (The detailed analysis has not been provided due to space limitations. Reader may refer [2] for the technical details of the combinatorial approach).\newline 
Now, the authors note that the sampling of spanning trees , according to the exponential family distribution, constructed above, follows many rules of $\lambda$-random trees.\ So, we need to define what is a $\lambda$-random tree. Given, values $\lambda_{e}\geq 0$, for all edges $e\in E$, a $\lambda$-random tree $T$ of the graph $G$ is a tree chosen from the set of all spanning trees of $G$, with a probability which is proportional to the product of the values of all the $\lambda_{e}$\ s, for the edges $e$ in the tree $T$.\ If we do a careful observation of the method of sampling the spanning trees of $G$ according to the exponential family distribution, we could see that the tree $T$ sampled is indeed a $\lambda$-random tree for $\lambda_{e}=e^{\gamma_{e}}$.The main idea for constructing a $\lambda$-random tree is by an iterative approach, in which we order all the edges of $G$ in an arbitrary fashion and then deciding probabilistically, whether to add a given edge to the tree or discard it. More precisly, if we know one edge that is in the tree for our initial assumption and all the values, $\lambda_{e}$\ s, then the probability, $p_{j}$ for an edge $e_{j}$ to be in the tree can be obtained by contracting all edges that have already been decided to be added to the tree and deleting all the other edges, that have already been decided to be deleted.\ Hence, we need to calculate the probability that some edge is in a $\lambda$-random tree, given all the $\lambda_{e}$\ s to find the $p_{j}$\ s to construct the tree and efficient methods for this calculation are well known.\ Now, we will define the thinness property of a tree, $T$ and use a known theorem for the concentration bound of the $\lambda$-random trees to show that if we sample $O(\log n)$ independent trees according to the distribution, $\tilde{p}(.)$, then we have high chances that the tree minimizing the cost, among all the sampled ones, is thin and cannot exceed two times the optimal solution cost of the Held-Karp bound. \newline \newline 
\textit{Definition 1}: A tree $T$ is $(\alpha ,s)$ thin if for each possible set $U\subset V$, we have $|T \cap \delta(U)|\leq \alpha z^{*}(\delta (U))$ and the cost of it does not exceed, $s.OPT_{HK}$. \newline \newline 
Moreover, for each edge $e\in E$ and a tree $T$ consider that, we have a random variable, $X_{e}$, such that $X_{e}=1$, when $e\in T$ and is 0, otherwise. If, for some $C \subset E$, we have $X(C)=\sum_{e\in C}X_{e}$, then the authors use the fact (it is also a theorem) that $Pr[X(C)\geq (1+\delta)\mathbb{E}[X(C)]]\leq \left(\frac{e^{\delta}}{(1+\delta)^{1+\delta}}\right)^{\mathbb{E}[X(C)]}$.\ Now, we define the following lemma. \newline \newline 
\textit{Lemma 1}: Consider, we sample a spanning tree $T$ according to the exponential family distribution, $\tilde{p}(.)$, of a graph $G=(V,E)$, where $|V|\geq 5$, then for any set, $U\subset V$, we have $Pr[|T \cap \delta(U)|> \beta z^{*}\delta(U)]\leq n^{-2.5z^{*}\delta(U)}$, for some $\beta = \frac{4 \log n}{\log \log n}$ and $\epsilon=0.2$. \newline \newline 
\textit{Proof}: Since, we are approximating the vector, $z^{*}$ with the vector $\tilde{z}$ and we are sampling the tree, $T$ according to $\tilde{p}(.)$, hence, as mentioned earlier, we have $\mathbb{E}[|T\cap \delta(U)|]=\tilde{z}(\delta(U))\leq (1+\epsilon)z^{*}\delta(U)$ for $U\subset V$.\ Let us take $1+\delta=\beta. \frac{z^{*}\delta(U)}{\tilde{z}(\delta(U))}\geq \frac{\beta}{1+\epsilon}$. We also, have $\beta. z^{*}\delta(U)\geq \frac{\beta. \mathbb{E}[|T\cap \delta(U)|]}{1+\epsilon}$ and using, $1+\delta \geq \frac{\beta}{1+\epsilon}$, we have 
\begin{equation*}
Pr[|T\cap \delta(U)|>\beta.z^{*}\delta(U)]\leq Pr[|T\cap \delta(U)|>(1+\delta).\mathbb{E}[|T\cap \delta(U)|]\ ]
\end{equation*}
\begin{equation*}
\hspace*{-2.7cm}\leq \left(\frac{e^{\delta}}{(1+\delta)^{1+\delta}}\right)^{\mathbb{E}[|T\cap \delta(U)|]}=\left(\frac{e^{\delta}}{(1+\delta)^{1+\delta}}\right)^{\tilde{z}(\delta(U))}
\end{equation*}
\begin{equation*}
\leq \left(\frac{e^{1+\delta}}{(1+\delta)^{1+\delta}}\right)^{\tilde{z}(\delta(U))}=\left(\frac{e}{1+\delta}\right)^{(1+\delta)\tilde{z}(\delta(U))}=\ \left(\frac{e}{1+\delta}\right)^{\beta z^{*}(\delta(U))}
\end{equation*}
\begin{equation*}
\hspace*{-3.95cm}\leq \left[\left(\frac{e(1+\epsilon)}{\beta}\right)^{\beta}\right]^{z^{*}(\delta(U))}\leq n^{-2.5z^{*}(\delta(U))}
\end{equation*}
which, could be obtained by basic algebra and taking $\epsilon=0.2$. \newline \newline 
Now, we are in a position, to state one of the most important theorems of the section, below. \newline \newline 
\textit{Theorem 2}: Consider, we have a collection of $\lceil 2\log n\rceil$, independently drawn sample trees of $G=(V,E)$, according to the distribution $\tilde{p}(.)$ and let, $T^{*}$ be the tree, with the minimum cost among all of them.\ Then, $T^{*}$ is $(4\frac{\log n}{\log \log n},2)$ thin with a high probability, for $n\geq 5$ and some $\epsilon = 0.2$. \newline \newline 
\textit{Proof}: Let us have the trees, $T_{1},T_{2},...,T_{\lceil 2\log n \rceil}$ and for any, $j\in [\lceil 2\log n \rceil]$ and a cut, $\delta(U)$, we know, from the previous lemma that
\begin{equation*}
Pr[|T_{j} \cap \delta(U)|> \beta z^{*}\delta(U)]\leq n^{-2.5z^{*}\delta(U)}
\end{equation*}
Also, due to Karger, we know that, there can be at most $n^{2l}$ cuts whose sizes do not exceed $l$ times the value of the minimum cut size, hence, we have at most $n^{l}$ cuts $\delta(U)$, such that $z^{*}(\delta(U))\leq l\left(1-\frac{1}{n}\right)$, for $l\geq 2$.\ Hence, the probability that there exists some cut, $\delta(U)$, which violates the $\beta$ thin-ness of some tree $T_{j}$ is at most, $\sum_{i=3}^{\infty}n^{i}.n^{-2.5(i-1)(1-1/n)} \leq \frac{1}{n-1}$, for $n\geq 5$, which is a very low value.\ Again, $\mathbb{E}[c(T_{j})]\leq \sum_{e\in E}\tilde{z_{e}}\leq \sum_{e\in E}(1+\epsilon)z_{e}^{*}\leq (1+\epsilon)OPT_{HK}$.\ By Markov's inequality, we have $Pr[c(T_{j})>2OPT_{HK}]\leq \frac{(1+\epsilon)OPT_{HK}}{2OPT_{HK}}=\frac{(1+\epsilon)}{2}$. Since, we drew $2\log n $ independent samples, hence the probability that the cost of the minimum tree exceeds $2OPT_{HK}$, cannot exceed $\left(\frac{(1+\epsilon)}{2}\right)^{2\log n}<\frac{1}{n}$.\ Hence, $Pr[c(T^{*})\leq 2OPT_{HK}]> 1-\frac{1}{n}$, which is a very high value. \newline \newline 
So, we finally have our thin spanning tree, $T^{*}$ and we will finally, convert it to an Eulerian walk and finally convert it to a Hamiltonian cycle, by shortcutting, of no greater cost.\ At first, we will modify our tree, $T^{*}$ to a directed tree, $T_{D}^{*}$, then construct upper and lower circulation capacity functions according to the rules of Hoffman's circulation theorem and formulate the problem as a minimum cost circulation problem and finally show, that the resulting Hamiltonian cycle, has not much higher cost.\newline \newline 
Let us re-orient each undirected edge, $\{u,v\}$ in the tree $T^{*}$ by looking the corresponding directed edges, $(u,v)$ and $(v,u)$, if they are present in the directed edge set $A$ (at least one must be present)
and then choosing the one with minimum cost. Also, due to the undirected nature of the cost function, $c$ we have $c(T_{D}^{*})=c(T^{*})$. For, each edge $a$, the lower capacity function, $l(a)$ could be designed as $l(a)=1$ if $a\in T_{D}^{*}$, otherwise it is 0.\ Similarly, the upper capacity function, $u(a)$ is $(1+2\alpha x_{a}^{*})$, if $a\in T_{D}^{*}$, else it is $2\alpha x_{a}^{*}$.\ The optimum solution of the minimum cost circulation problem, say $f^{*}$, which could be computed in polynomial time, corresponds to a directed multigraph, say $G_{f}$, which has the directed tree, $T_{D}^{*}$ and $G_{f}$ is an Eulerian directed multigraph, which could be shortcutted to obtain a Hamiltonian cycle of cost not exceeding $c(f^{*})$.\ We have, $c(f^{*})\leq c(u)$ and since, $u(a)=(1+2\alpha x_{a}^{*})$, if the edge $a$ belongs to the directed tree, $T_{D}^{*}$, thus $c(u)=c(T_{D}^{*})+2\alpha c(x^{*})$.\ Again, since $T_{D}^{*}$ is $(\alpha,s)$ thin, hence we have $c(T_{D}^{*})\leq s.OPT_{HK}=s.c(x^{*})$. So, $c(u)\leq (2\alpha+s)c(x^{*})$.\ These could be summarized as a theorem, as stated below. \newline \newline 
\textit{Theorem 3}:\ Given an $(\alpha,s)$ thin spanning tree, $T^{*}$, corresponding to the solution of the LP relaxation, $x^{*}$, we can always find a Hamiltonian cycle efficiently of cost not exceeding $(2\alpha+s)OPT_{HK}$, where $OPT_{HK}=c(x^{*})$ is the optimal cost.\newline \newline 
Since, we already obtained a $(\frac{4\log n}{\log \log n},2)$ thin spanning tree with high probability, hence by the above theorem, a Hamiltonian cycle could be found out efficiently of cost at most $(2+\frac{8\log n}{\log \log n})=O(\frac{\log n}{\log \log n})$ of the optimal value.\ The overall algorithm could be summarized as below: \newline \newline
\textit{Input}: A set of $n$ vertices and a positive real cost function, satisfying the triangle inequality \newline \newline 
1.\ Solve Held-Karp, find the optimal solution, $x^{*}$ and construct scaled down symmetric vector, $z^{*}$ \newline 
2.\ Sample $O(\log n)$ spanning trees from the exponential family distribution, $\tilde{p}(.)$ and find the tree with the minimum cost. \newline 
3.\ Orient the edges of the tree to form a directed graph, find a minimum cost circulation that contains the tree and shortcut this multigraph to form a Hamiltonian cycle. Output the cycle.

\end{document}